\documentclass[epsfig,12pt]{article}
\usepackage{epsfig}
\usepackage{graphicx}
\usepackage{array}
\usepackage{color}
\usepackage{bm}
\usepackage{cite}
\usepackage{geometry}
\usepackage{latexsym}
 \usepackage{amsmath, amssymb, amscd, xypic, graphicx}

\newcommand{\beq}{\begin{equation}}   
\newcommand{\eeq}{\end{equation}}
\newcommand{\beqn}{\begin{eqnarray}}   
\newcommand{\eeqn}{\end{eqnarray}}

\newcommand{\pt}{\partial}

\def\ntwo{${\mathcal N}=2\;$}

\newcommand*\xbar[1]{%
 \kern0.5ex%
  \hbox{%
   \kern0.2ex%
      \vbox{%
      \hrule height 0.5pt % The actual bar
      \kern0.5ex%         % Distance between bar and symbol
      \hbox{%
        \kern-0.1em%      % Shortening on the left side
        \ensuremath{#1}%
        \kern-0.1em%      % Shortening on the right side
      }%
    }%
  }%
}

\newcommand{\gsim}{\lower.7ex\hbox{$
\;\stackrel{\textstyle>}{\sim}\;$}}
\newcommand{\lsim}{\lower.7ex\hbox{$
\;\stackrel{\textstyle<}{\sim}\;$}}
\begin{document}

\begin{titlepage}

\begin{flushright}
FTPI-MINN-17/10, UMN-TH-3626/17
\end{flushright}

\vspace{5mm}

\begin{center}
{  \Large \bf  
Fradkin-Shenker Continuity and \\[2mm] ``Instead-of-Confinement'' Phase}
 
\vspace{5mm}
%\vspace{1mm}

{\large \bf   M.~Shifman$^{\,a}$ and \bf A.~Yung$^{\,\,a,b,c}$}
\end {center}

\begin{center}

$^a${\it  William I. Fine Theoretical Physics Institute,
University of Minnesota,
Minneapolis, MN 55455}\\
%[1mm]
$^{b}${\it National Research Center ``Kurchatov Institute'', 
Petersburg Nuclear Physics Institute, Gatchina, St. Petersburg
188300, Russia}\\
$^{c}${\it  St. Petersburg State University,
 Universitetskaya nab., St. Petersburg 199034, Russia}
\end{center}

\vspace{1cm}

%\vspace{6mm}

\begin{center}
{\large\bf Abstract}
\end{center}

In 1979 Fradkin and Shenker observed \cite{Frad} that if one considers a Yang-Mills theory fully Higgsed
by virtue of scalar fields in the fundamental representation of the ${\rm SU}(N)$ gauge group there is 
no phase transition in passing from the Higgs regime (weak coupling) to the ``QCD confinement" 
regime at strong coupling. The above two regimes are continuously connected. We combine this observations with  
lessons from 
supersymmetric gauge theories which show that the Higgs phase is continuously connected
to what is called ``instead-of-confinement'' phase rather than the phase with quark confinement.
In the ``instead-of-confinement'' phase monopoles are confined and play a role of ``constituent''
quarks inside hadrons. In contrast,  the Seiberg-Witten  phase of quark confinement is not
analytically connected to the Higgs phase.

We propose dedicated lattice studies of Yang-Mills theories with scalar quarks.

\vspace{2cm}

\end{titlepage}

\section{Introduction}
\label{1}

Yang-Mills theory and its QCD versions are strongly coupled; their large distance dynamics  defies analytic solutions for four decays. One of possible strategies of obtaining certain insights in the  strong coupling regime is to introduce an extra 
adjustable parameter which one could continuously vary from weak coupling to strong coupling. 

In this paper we re-visit the so-called Fradkin-Shenker continuity \cite{Frad} --
a continuity between the strong and weak coupling (Higgsed) regimes in Yang-Mills theory in four dimensions. 
We will confront this picture with lessons from recent supersymmetry-based results \cite{SYrev,16}, more exactly, 
some qualitative aspects ensuing from these results. To this end non-supersymmetric Yang-Mills theory will be Higgsed in precisely the same way as the supersymmetric theory from which we draw our inspiration. 

Needless to say, in non-supersymmetric Yang-Mills models powerful tools for obtaining exact results based on holomorphy are lost and
we cannot make fully quantitative predictions. However, we expect the overall qualitative picture emerging in
supersymmetric QCD to be preserved when supersymmetry is lifted. This will allow us to get a better 
understanding of QCD-like theories at strong coupling. This understanding turns out to be rather surprising. 

Fradkin-Shenker picture motivated by  lattice theory is as follows:  if one introduces Higgs fields 
in the fundamental representation in such a way that all gauge fields are Higgsed, 
and then studies the theory as a function of the vacuum expectation value $v^2$, 
then at positive  $v^2\gg \Lambda^2$ ($\Lambda$ 
is the dynamical scale of the theory) one has weak coupling regime, while diminishing $v^2$ and crossing
the line $v^2\lsim \Lambda^2$ one attains a strong coupling regime similar to ``QCD confinement," i.e. confinement with dynamical light (scalar) quarks. Then further downward evolution of $v^2$ (i.e., making $v^2$ vanishing or negative 
with $|v^2|\lsim \Lambda^2$) will change nothing. No phase transition is supposed to occur; there are no two 
distinct phases -- Higgs vs. confinement -- because there is no order parameter to distinguish them. One always has perimeter law rather than area law. Fradkin and Shenker suggested a picture of a single Higgs-confinement phase, with the perimeter law for the Wilson loop on both sides of the weak-strong coupling regimes. 

Let us explain what is meant above by ``QCD confinement." 
With light scalar quarks in the fundamental representation even at strong coupling there is no confinement in the strict 
sense of this word: large contours are screened and no Wilson area law emerges. The quark pair creation breaks
{\em bona fide} strings. The strings are not infinitely long. Nevertheless we can guess 
that they are there because if quarks have flavor quantum numbers, in addition to color, we do 
not observe these quantum numbers in the mesonic spectrum,
only those which are inherent to quark-antiquark bound pairs are asymptotic states.\footnote{Baryons are not seen at  
 large $N$.} Even if the quarks are not light, strictly speaking the area law is recovered only in the limit
 $m\to\infty$, where $m$ is the quark mass term. 
 
 It is universally believed that replacing the light scalar quarks by light Dirac quarks in the same fundamental representation
 does not change the essential aspects of the above picture. 
 
 If solution of supersymmetric QCD can be used as an indication for non-supersym\-metric version, this general belief may not be correct. ``QCD confinement" regimes for theories with scalar and spinor quarks could be different, as is clearly different the confinement regime in pure Yang-Mills in which strings are unbreakable and the area law applies. 

In this paper we compare the Fradkin-Shenker continuity with the picture obtained 
in \ntwo supersymmetric QCD (SQCD). First we briefly review the Fradkin-Shenker description of  
non-supersymmetric version of Yang-Mills theory endowed with the Higgs sector described 
in detail in Sec. \ref{TWO}. This theory has  the action similar to the bosonic part of the action of
\ntwo SQCD. 

In the Higgs regime (positive $v^2$) at weak coupling ($v^2\gg\Lambda^2$) color charges 
are screened through Higgsing. The residual  flavor symmetry of the model is
${\rm SU}(N)_{\rm global}$. There is no long-range forces in the theory and no bound states. Asymptotic states can be read off  from the Lagrangian. All gauge bosons become massive and are in the adjoint of the ${\rm SU}(N)_{\rm global}$. The scalar quarks which used to be  {\em fundamentals} of  the global group before Higgsing become ${\rm SU}(N)_{\rm global}$ {\em adjoints} after Higgsing. 

As we diminish $v^2$ it approaches $v^2 \lsim \Lambda^2$, and may  cross zero and  become negative. Let us assume that $v^2$  freezes at a value below $\Lambda^2$. Then we are in the strong coupling regime. 

At this point we use the  picture obtained in \ntwo SQCD. Exact results obtained from
supersymmetry show that at strong coupling this theory does not go into the confinement phase. Instead
it ends up in a novel phase, which we called ``instead-of-confinement''  \cite{SYdual,15}, see
\cite{16} for a review.  In this phase there is no confinement of quarks --
 quarks and gluons screened at weak coupling evolve at strong coupling into
monopole-antimonopole pairs confined by  strings. The role of constituent quarks inside
hadrons is played by monopoles which carry appropriate flavor quantum numbers.
 This phase is qualitatively
rather similar to what we observe in real-world QCD. At the same time, we show that the Seiberg-Witten 
phase of quark confinement is not analytically connected to the 
Higgs-``instead-of-confinement'' phase in SQCD.

We believe that qualitatively the above picture may be generic rather than specific to SQCD. 
Supersymmetry just helps us to solve the theory. We suggest that {\em non-supersymmetric} scalar
QCD and \ntwo supersymmetric QCD are  in the same universality class. In particular, we expect that 
non-supersymmetric scalar QCD continuously evolve from the Higgs phase at large $v^2$ to 
the ``instead-of-confinement'' phase rather than to the quark confinement phase at small $v^2$. 
The discussion below may be interpreted as an example that light scalar quarks vs. light Dirac quarks
coupled to Yang-Mills at strong coupling are not as close theories as they were thought of.

The presence of a parameter which interpolates between weak and strong coupling regimes quite often 
is considered as a tool to reveal physics at strong coupling.
 One may hope that if it is independently known that
there is no phase transitions on the way, one can carry out analytic analysis at weak coupling 
and then extrapolate -- qualitatively or semi-quantitatively --  the
picture thus obtained to strong coupling.
However, one must clearly understand what particular theory is obtained after the analytic 
continuation. 

We stress that although there is no phase transition between weak and strong coupling
regimes in \ntwo SQCD (in accordance with the Fradkin-Shenker continuity) there is a crossover. 
In particular, light ``fundamental'' fields of the dual theory describing  physics at small $ v^2 $ -- 
``dual quarks''  -- are represented by  solitonic states in terms of quarks and gluons of the 
original theory. In this sense the hope that if
there is no phase transition on the way then one can straightforwardly extrapolate  the weak coupling
picture to strong coupling may turn out an illusion.

The paper is organized as follows. In Sec.~\ref{TWO} we review the scalar QCD at weak coupling
and in Sec.~\ref{largeN} comment on the large $N$ limit.
In Sec.~\ref{N=2QCD} we discuss the picture obtained from \ntwo supersymmetric QCD and in 
Sec.~\ref{SWconfinement} we comment on the Seiberg-Witten confinement phase. Sec.~\ref{conclusions}
contains our conclusions.

\section{The basic set up}
\label{TWO}

The basic non-supersymmetric model  has the form
\beq
{\cal L}_0 = -\frac{1}{4g^2} F_{\mu\nu}^p  F^{\mu\nu\, p}  -\frac{1}{4g^2_1} F_{\mu\nu}  F^{\mu\nu} 
+ \sum_A\left| {D_\mu} \phi_A^k \right|^2 - V(\phi)\,,
\label{fr1}
\eeq
where $p$ is the adjoint index of the color ${\rm SU}(N)$, ($p=1,2,... , N^2-1$), $k$ is the fundamental index of the color ${\rm SU}(N)$, $k=1,2,... , N$,  and the subscript $A$ labels the fields of the scalar sector, $A= 1,2,... , N$. 
The gauge symmetry in \eqref{fr1} is ${\rm U}(N)={\rm U}(1)\times {\rm SU}(N)$. The covariant derivative is
\beq
D_{\mu}=\pt_{\mu} -\frac{i}{2}A_{\mu} -iT^pA_{\mu}^p\,,
\eeq
where $A_{\mu}$ and $A_{\mu}^p$ are U(1) and SU$(N)$ gauge potentials, while $T^p$ are generators of 
the color SU$(N)$. To make our discussion simpler we assume that U(1) and SU$(N)$ coupling constants
$g^2_1$ and $g^2$ are related 
\beq
g^2= \sqrt{\frac{N}{2}}\,g_1^2
\eeq
(although this assumption could be lifted).
The scalar sector is assumed to have a global (flavor) ${\rm SU}(N)$ symmetry labeled by 
the index $A$ (see Eq. (\ref{fr1}))
so that the potential $V(\phi )$ must be chosen appropriately. 
Our choice will be motivated by supersymmetric field theory (see e.g. \cite{SYrev}), 
\beq
V= \lambda_1 \sum_p\left(\sum_A \bar\phi^A T^p \phi_A\right)^2+\lambda_2 \left(\sum_A \bar\phi^A  \phi_A - Nv^2\right)^2 \,.
\label{fr2}
\eeq
Here  $\lambda_{1,2}$ are constants. If $\lambda_{1,2} \sim g^2$,
then (\ref{fr2}) represents a (somewhat reduced)  bosonic sector of the supersymmetric theory.
For what follows it is convenient to introduce $N\times N$ matrix $\Phi = \{ \phi^i_A\}$ which is constructed of $N$ columns
\beq
\Phi \leftrightarrow \{ \phi^k_1,\, \phi^k_2, \, ... , \phi^k_N\}
\label{fr3}
\eeq
where the subscript marks flavor while the superscript $k$ refers to color (in  the fundamental representation).
In this notation
\beqn
{\cal L}_0 &=& -\frac{1}{4} F_{\mu\nu}^p  F^{\mu\nu,\, p} -\frac{1}{4} F_{\mu\nu}  F^{\mu\nu}+ {\rm Tr}\, \left( {D_\mu} \Phi \right)^\dagger
\left( {D^\mu} \Phi \right) -V(\phi)\,,\nonumber\\[3mm]
V&=& \lambda_1 \sum_p\left( {\rm Tr}\, \Phi^\dagger T^p \Phi\right)^2+\lambda_2 \left[ {\rm Tr }\left( \Phi^\dagger  \Phi - Nv^2\right) \right]^2 \,.
\label{fr4}
\eeqn

Now, it is perfectly clear that if $v^2$ is positive $\Phi$ develops the diagonal expectation value of the form
\beq
\Phi_{\rm vac} =\left(\begin{array}{cccc}
v&0&...&0\\[1mm]
0&v&...&0\\[1mm]
...&...&...& ...\\[1mm]
0&0&...&v
\label{fr5}
\end{array}
\right)
\eeq
 (up to   irrelevant gauge transformations). 
%The parameter $v$ can be chosen to be positive, using the ${\rm U}(N)$ 
% gauge freedom. 
This form of the vacuum expectation values (VEVs) was suggested long ago \cite{Bardakci}.
 
  All $N^2$ gauge bosons of the original ${\rm U}(N)$ gauge theory are Higgsed, acquiring one and the same 
  mass $M_W=gv$. All $N^2$ real physical Higgses acquire masses, $N^2$
 ``phases" are eaten up by gauge bosons. 

Equation (\ref{fr5}) shows that in the Higgs regime both the local gauge and global flavor groups are spontaneously
broken, but the diagonal ${\rm SU}(N)_{C+F}$ survives as the {\em exact global} symmetry of the model. 
The $M_W=gv$ vector bosons form the singlet and adjoint representations of the above global group, so we will call them $W$ bosons.
They can be defined in the gauge invariant form  as follows:
\beq
g\left( W_\mu\right)^A_B = i\,c \, \left( \bar\phi^A_i \,  \! \stackrel{\leftrightarrow}{D_\mu} \phi^i_B \right) .
 \label{fr7}
\eeq
where $c$ is a normalization factor.
Their effective low-energy self-interaction of dimension three and four is similar to that of gluons 
in the microscopic Lagrangian. The physical Higgs bosons
\beq
H_A^B = \tilde{c} ({\bar\phi}_i^B \phi^i_A -v^2\delta_{A}^{B}) 
 \label{fr8}
\eeq
 form a singlet and adjoint representations with respect to the exact global symmetry ($N^2$ real fields altogether).
 Here $\tilde c$ is another normalization factor.
  The interaction vertices between the physical Higgses in the adjoint representation of 
${\rm SU}(N)_{C+F}$ and the massive $W$ bosons are proportional to
 Tr$(\partial^\mu H^\dagger ) W_\mu H+ ...\,$. Overall, the effective low-energy theory in the
 Higgs regime looks like Yang-Mills with massive gauge bosons. Of course, it is not meant to calculate loops. 

At large $v$ the spectrum of the theory consists of vector and scalar particles specified above. 
Of course, we can combine a number of these particles to create  states with arbitrary spins, angular momentum, and  belonging 
to various representations of ${\rm SU}(N)_{C+F}$. They will not be bound states, however, because
bound states do not form  for arbitrarily weak coupling constant unless there are long range forces, absent in our case.
 
One type of  non-perturbative objects discussed at weak coupling (in the Euclidean space)  are instantons \cite{BPST,hooft}. 
Note that the instanton action as well as the mass of the confined monopole-antimonopole pair 
(see Sec.~\ref{N=2QCD})
scale as $N$ at large $N$. The Euclidean instanton is a reflection of the existence of non-perturbative quasiclassical objects in the Minkowski space.\footnote{ {\em A remark about renormalons in passing}: \hspace{1mm} Occurrence of infrared renormalons in Yang-Mills theories \cite{renormalon} (for a recent review \cite{ural}) is a formal indicator for $N$-independent nonperturbative effects. In the Higgs regime described above infrared 
renormalons are absent. Indeed, since the vector bosons of the theory are massive (and $M^2\gg\Lambda^2$), the factorial growth of the series
$$
\sum_n a_n \left(\alpha(Q^2) \right)^n\,,\quad a_n \sim n! \,\,\,\, \mbox{at}\,\, n\gg 1
$$
stops at 
$$
n_* =2\log\frac{Q^2}{M^2}\,,
$$
and at $n> n_*$  the factorial divergence of the coefficients disappears, and so do singularities in the Borel plane.
The series become formally summable, see Sect. 6 in \cite{ural}. }
 
\section{Large-\boldmath{$N$} limit}
\label{largeN}
 
 The model introduced in Sec. \ref{TWO} has a significant advantage over many other models in the Higgs regime.
 Namely, it allows one to consider a smooth large-$N$ limit, analogous to the 
  't Hooft limit. More exactly, since the number of flavors in the given model is equal to the
  number of colors, here we deal with the Veneziano limit \cite{vene}.

 It is clear that the dynamical scale parameter $\Lambda$
 which is kept fixed as $N\to \infty$ is now replaced by the mass $M_W = gv$ of the vector bosons. Therefore the scaling law is
 \beq
 N\to\infty\,,\quad g^2 N \,\,\mbox{fixed}\,,\quad g^2v^2 \,\,\mbox{fixed}\,,\quad v^2 \sim N\,.
 \label{scla}
 \eeq 
 Under this scaling both the BPST part of the instanton action and the 't Hooft part scale as $N$, and so does
 the sphaleron mass. Thus, we must be careful with interchanging limits of large energies and large  $N$. We may keep in mind a large value of $N$ but $N\neq\infty$. In principle, $N$ does not have to be large in what follows.

\section{Microscopic picture from supersymmetric QCD}
\label{N=2QCD}

QCD with $N$ flavors of scalar quarks discussed above is not yet solved at strong coupling.
Our goal now is to compare the Fradkin-Shenker continuity in the model at hand with the results
obtained from our previous considerations of ${\cal N}=2$ supersymmetric models, in which
the microscopic picture of the underlying phenomena can be revealed, see  \cite{SYrev,16} for reviews. 
The regime which may be closest to what we see in real-world QCD  at strong coupling 
has been studied 
in \cite{SYdual,15} where it is referred to
as ``instead-of-confinement." Why ``instead" will become clear shortly. 

We will narrow down a class of theories considered in \cite{SYdual,16} to a subclass with the  bosonic sector coinciding with 
that of Sec. \ref{TWO}. Namely, we will discuss theories with the U$(N)$ gauge group which have equal 
numbers of colors and quark hypermultiplets, i.e. 
$$N_f=N\,,$$  in the so-called quark vacuum. In this vacuum at large $v^2$  all $N$ squark flavors condense through the potential similar to (\ref{fr2}), so that color-flavor locking takes place. The Fayet-Iliopoulos
$D$-term is {\em not } introduced. Instead,
we introduce equal mass terms $m$ to all flavors. Each flavor consists of  complex scalar superfields 
$q^{k A}$ and $\tilde{q}_{k A}$
where $A$ is the flavor index, while $k$ is the color index (in the fundamental representation); both run from 1 to $N$. Each complex scalar field has a Weyl superpartner. All quarks have the same mass $m$.

In addition, a mass term $\mu$ is introduced for the adjoint scalar superfield,  generally speaking breaking ${\cal N}=2$ down to ${\cal N}=1$. However, the deformation superpotential 
\beq
{\mathcal W}_{{\rm def}}=
  \mu\,{\rm Tr}\,\left(\frac12\, {\mathcal A} + T^a\, {\mathcal A}^a\right)^2, \qquad 
\label{msuperpotbr}
\eeq
does not break \ntwo supersymmetry in the small-$\mu$ limit,  see \cite{HSZ,VY,SYmon}. The fields  ${\mathcal A}$ and ${\mathcal A}^a$ 
in Eq.~(\ref{msuperpotbr}) are  chiral superfields, the ${\mathcal N}=2$
superpartners of the U(1) and SU($N$) gauge bosons.  We will treat $\mu$ as a free parameter which can be taken arbitrarily small. At small $\mu$ the deformatiom superpotential reduces to the 
Fayet-Iliopoulos $F$-term.

\subsection{Weak coupling: Higgs regime}

At weak coupling the gauge group is fully Higgsed, the squark VEVs are given by \eqref{fr5} 
with the identification 
\beq
q^{kA}=\bar{\tilde{q}}^{kA} =\frac{1}{\sqrt{2}}\phi^{kA}.
\label{a12}
\eeq
The role of $\lambda_{1,2}$ in Eq. (\ref{fr4}) is played by $g^2$,
\beq
\lambda_1= \frac{g^2}{2}, \qquad \lambda_2= \frac{g^2}{8}\,\sqrt{\frac{2}{N}},
\eeq
while 
\beq
v^2 = | \mu \,m |\, ,
\eeq
where $m$ is a quark mass. In SQCD $v^2$ is always positive.\footnote{Generally speaking the parameters $\mu$ and $m$ may have phases. The {\em ansatz} (\ref{a12}) assumes that this phases are rotated away, which one can always do.} In non-supersymmetric theory 
of Sec. \ref{TWO} the value of $v^2$ can be both positive and negative. 

The unbroken global symmetry of this model is  
\beq
  {\rm SU}(N)_{C+F}.
\label{c+f}
\eeq
The global $ {\rm SU}(N)_{C+F}$ symmetry is responsible for the formation of the non-Abelian strings 
 \cite{HT1,ABEKY,SYmon,HT2} with the world-sheet theory described by the ${\cal N}=2$ ${\rm CP}(N-1)$ model. In the limit $\mu\to 0$, to the first order in $\mu$, these strings are BPS saturated and their tension is 
 \beq
T =2\pi v^2\, .
\label{ten}
\eeq
 Note that at large $N$ the scaling of $v^2$ is given by Eq. (\ref{scla}).

 \vspace{1mm}
 
 What is the spectrum of this theory at weak coupling?
 
 \vspace{1mm}
 
 All states come in 
representations of the unbroken global
 group (\ref{c+f}), namely, in the singlet and adjoint representations
of SU$(N)_{C+F}$, as was mentioned in Sec. \ref{TWO}. The perturbative states are screened quarks  
and gauge bosons, supermultiplets containing $H_A^B$ and   $W_A^B$.  There are also non-perturbative states,
see \cite{SYrev} for a detailed review.

\begin{figure}
\centerline{\includegraphics[width=7cm]{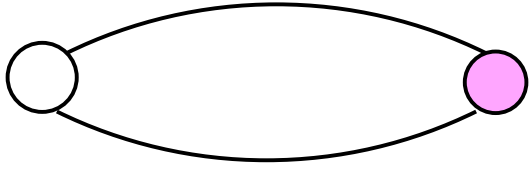}}
\caption{\small Meson formed by a monopole-antimonopole pair connected by two strings.
Open and closed circles denote the monopole and antimonopole, respectively.}
\label{figmmeson}
\end{figure}

Since the large-$v^2$ the theory  is in the Higgs regime for squarks,  non-Abelian strings confine monopoles.
In fact in the U$(N)$ gauge theories non-Abelian strings are stable, a string cannot terminate on a monopole.
Instead, the monopoles become junctions of two distinct non-Abelian strings.
This leads to occurrence of mesons formed by monopole-antimonopole pairs confined by
non-Abelian strings \cite{SYrev}, see Fig.~\ref{figmmeson}.

The flavor quantum numbers of stringy  monopole-antimonopole mesons were studied in 
\cite{SYtorkink} in the framework of an appropriate two-dimensional   ${\rm CP}(N-1)$  model which describes 
world-sheet dynamics of the non-Abelian strings \cite{HT1,ABEKY,SYmon,HT2}. In particular, 
confined monopoles 
are seen as kinks in this world-sheet theory. The kinks belong to the fundamental representation of 
the global SU$(N)_{C+F}$ \cite{W79}. Therefore, 
if two strings in Fig.~\ref{figmmeson} are ``neighboring''
 each meson is in the two-index representation $M_A^B$
of the flavor group, where the flavor indices are 
$A,B =1,..., N$. It splits into the singlet and  adjoint representations of 
the global unbroken group SU$(N)_{C+F}$. 

At weak coupling the  stringy mesons are heavy. Their masses are given by the product of the string tension 
times the length of the string $M\sim v^2 L$. The string length is bounded from below by the string width,
which is determined by masses of the bulk excitations forming the string, $L\gtrsim 1/gv$. This gives
\beq
M\sim \frac{v}{g}.
\label{mesonmass}
\eeq
The stringy mesons are highly unstable and decay into
perturbative (screened) quarks and gluons  $H^B_A$ and $W^B_A$ which are lighter at large $v$, with masses $\sim M_W=gv$.
The stringy mesons may be related to the sphaleron solution  \cite{sph}.

\subsection{Strong coupling: ``instead-of-confinement'' regime}

Now let us pass to strong coupling, i.e. the domain of small $v^2$. 
In non-supersymmet\-ric theories such as scalar QCD \eqref{fr4} this step cannot be carried out 
analytically.
This is the point where supersymmetry becomes important. More exactly,
we exploit the exact Seiberg-Witten solution  on the Coulomb branch \cite{SW1} in our theory. We start at large $v^2 \sim |\mu m |$ (in the large-$m$ limit; all quark masses are equal) and then
go to the equal mass small-$v^2$ limit  via the domain of large 
$\Delta m \sim \Delta m_{AB}\equiv (m_A-m_B)$ \cite{SYdual}.

Now let us have a closer look at the spectrum and other features of the strong coupling (small $v^2$) regime. We will deduce
these features from the dual theory. In accordance with the general Fradkin-Shenker continuity there is 
no phase transition on the way from weak to strong coupling, just a crossover. Therefore the global 
symmetry group of the dual theory is the same as in the original one, SU$(N)_{C+F}$, see
  Eq. (\ref{c+f}).

As was shown in \cite{SYdual,15} (see also \cite{16}) the domain of small $| v^2|$ can be described  
by a weakly coupled  dual theory\,\footnote{The ${\rm U}(1)^{N}$ gauge group for the dual theory was first 
obtained by Seiberg and Witten  \cite{SW1,SW2} in the monopole vacua, see also \cite{APS}.} with the
gauge group ${\rm U}(1)^{N}$. 
Needless to say, this theory is infrared free. The matter fields 
are light ``dual quarks"  $D^\gamma$. The index $\gamma=1,...,N$ numbers $D$'s. Altogether
there are $N$ such fields, all are singlets with respect to the global SU$(N)_{C+F}$, symmetry. The $\gamma_i$-th $D$
is charged with respect to the $\gamma_i$-th U(1) gauge group from the set ${\rm U}(1)^{N}$ mentioned above.

At strong coupling the dual quarks condense  \cite{SYdual}. The  ${\rm U}(1)^{N}$ dual gauge group is 
completely Higgsed. This leads to the formation of 
confining 
strings of the Abelian Abrikosov-Nielsen-Olesen (ANO)  type \cite{ANO} in each of gauge U(1) 
factor.\footnote{In fact these strings can still be understood as  vacua of the world sheet CP$(N-1)$ model
upon continuation to small $v^2$ \cite{SYtorkink}.} 
Since  $D^\gamma$ color charges are identical to those 
of diagonal squarks from $q^{kA}$ these ANO strings confine monopoles \cite{SYdual}. 

The fact that $D^\gamma$'s are singlets with respect to the global SU$(N)$ group 
shows that the ``dual quarks"  are not the quarks of the weakly coupled theory: 
the latter are in the adjoint representation of the
global SU$(N)_{C+F}$. 

Thus, the ``dual quarks" and  quarks are distinct states. 
At large $v^2$ ``dual quarks" are heavy solitonic states. However below the crossover
at small $v^2$ they become light and form the fundamental ``elementary" states $D^{\gamma}$ of the dual 
theory.

This raises the question: what  happens to quarks when we reduce $v^2$? 

They find themselves in the ``instead-of-confinement'' phase. The
Higgs-screened quarks $H_A^B$  and gauge bosons $W_A^B$
at small $v^2$  decay into the monopole-antimonopole 
pairs on the curves of marginal stability (the so-called wall crossing) \cite{SYdual,SYtorkink}. 
The latter are present in \ntwo supersymmetric QCD\cite{SW1,SW2,BF}.

 At small non-vanishing values of $v^2$ the
monopoles and antimonopoles produced in the ``decay" of  $H_A^B$  and $W_A^B$
cannot escape from each other
because they are confined. Therefore, the (screened) quarks and  gauge bosons 
evolve into stringy mesons similar to those which appear at weak coupling, see  Fig.~\ref{figmmeson},
namely monopole-antimonopole
pairs connected  by two confining strings \cite{SYdual,16}.

\begin{figure}
\epsfxsize=6cm
\centerline{\epsfbox{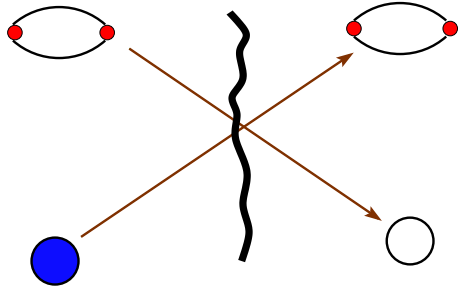}}
\caption{\small A schematic picture of the crossover from large to small 
$v^2$. Big closed circle denotes light quarks, while big open circle denotes
light ``dual quarks''. Monopole-antimonopole mesons are shown as in Fig.~\ref{figmmeson}.
Large values of $v^2$ are represented by the left-hand side, while small values by the right-hand side. }
\label{figlevelcross}
\end{figure}

The picture of the crossover is schematically shown in Fig.~\ref{figlevelcross}. The left- and 
right-hand sides of this figure correspond
to large and small values of $ v^2$ respectively. In \ntwo SQCD squarks are light at large $v^2 $.
They evolve into monopole-antimonopole stringy mesons at small $v^2$.
Moreover, heavy (unstable) monopole-antimonopole stringy mesons present at large $v^2$
become light at small $v^2$ and form ``fundamental'' charged matter
of the dual theory, namely ``dual quarks.'' Screened ``dual quarks'' (they are singlets with respect to
global the SU$(N)_{C+F}$) form glueballs.

To summarize at small $v^2$ the monopole-antimonopole stringy mesons in the two-index representations
of SU$(N)_{C+F}$  are descendants from (screened) quarks and gauge bosons of the original theory. 
We see that these mesons have ``correct''
adjoint or singlet quantum numbers with respect to the global group,
like mesons in the real-world QCD. Singlets are interpreted as glueballs.

Moreover, because these mesons are formed by strings they 
lay on Regge trajectories. 

Thus, the monopole-antimonopole mesons
of the instead-of-confinement phase are qualitatively similar to mesons of the real-world QCD. The role
of QCD constituent quarks is played by monopoles. If one could scatter external probes 
off such mesons one would see 
a gluon and scalar quarks in the deep inelastic limit.

Note also that baryons  are represented by  ``monopole necklaces'' formed by close string
configurations with $N$ confined monopoles attached \cite{SYrev}. However, what is usually 
called the baryonic U(1) symmetry is a part of the broken gauge group in our U$(N)$ SQCD. Therefore
baryons are unstable and decay into  mesons.

\section{Seiberg-Witten confinement: no continuity}
\label{SWconfinement}

Now let us comment on the Seiberg-Witten confinement. It occurs in the so-called
monopole vacua of \ntwo SQCD. In this vacua monopoles condense at non-zero $\mu$ and  the ANO strings
with color-electric fluxes
are formed. These strings confine  quarks \cite{SW1,SW2}.

Let us stress that this happens in the monopole vacuum which is a  different vacuum as compared to
the quark vacuum which we considered in the previous sections.  The quark vacuum is defined as the 
vacuum where at large quark masses $m$
($m\gg \Lambda$) all $N$ squark VEVs are large $v\sim \sqrt{|\mu m |}$, see \eqref{fr5}. In the monopole
vacuum squarks classically do not condense, while the monopole VEVs are much smaller, 
$\sim \sqrt{| \mu \Lambda|}$.
The monopole vacuum does not have a weak coupling description in terms of the original theory.

The monopole vacuum is not 
analytically connected to the quark vacuum. To see that this is the case
 note that the gluino condensate is identically zero
in the quark vacuum while in the monopole vacuum it is a non-trivial function of the quark mass $m$ 
known exactly \cite{Cachazo2}. Thus to jump from the quark vacuum at weak coupling to 
the monopole vacuum at
strong coupling we need to break analyticity, i.e. this passage involves a phase transition.

\section{If SQCD results shed light on QCD}

If the picture that was derived in the quark vacua of \ntwo SQSD shares qualitative features 
with nonsupersymetric QCD then we have to conclude that scalar and spinor 
QCD are not as close as was believed.
The former has an obvious weak coupling regime. At strong coupling mesons of the scalar QCD are
formed by  ``constituent scalar quarks'' which are color-magnetically charged monopoles  attached to each 
other by virtue of confining strings.
On the contrary, in the spinor QCD we expect the Seiberg-Witten confinement scenario when quarks 
themselves are  attached to each other by virtue of the confining strings. 

Moreover, in the nonsupersymmetric theory discussed in Sec. \ref{TWO}, in contradistinction to SQCD,
one can consider negative values of $v^2$ in Eq. (\ref{fr2}). At negative $v^2$ the scalar quarks do not condense. 
When we increase $|v^2|$ keeping $v^2$ negative, the mass of the scalar quark increases.
One can expect that its back reaction on the vacuum becomes suppressed. At a certain point the string
that develops between the heavy scalar quark and its antiquark should be the same  as that in pure Yang-Mills theory. 
Presumably, this would require a phase transition.

The distinction which our hypothesis outlines 
is hard to detect analytically. However, lattice studies could help 
verify it. We already know that spinor quarks undergo a chiral 
phase transition, which cannot happen in scalar QCD due to the absence of the chiral symmetry.

\section{Conclusions}
\label{conclusions}
 
In this paper we revisited the Fradkin-Shenker continuity. We assumed that lessons from
supersymmetry can prompt us how this  continuity  is implemented in scalar QCD. If so, we conclude that
the Fradkin-Shenker continuity connects the Higgs and 
the  ``instead-of-confinement'' regimes.
In the latter regime 
the quarks and gauge bosons screened at weak coupling evolve at strong coupling into
the monopole-antimonopole pairs confined by strings. 
 This regime is qualitatively
rather similar to what we observe in real-world QCD.
Supersymmetry also teaches us that the Seiberg-Witten phase of quark confinement is 
not analytically connected to the Higgs phase.  

Summarizing, we expect that 
  non-supersymmetric scalar QCD \eqref{fr4} continuously evolves from the Higgs regime at large $v^2$ to 
the ``instead-of-confinement'' regime at small $v^2$, rather than to the quark confinement phase. This passage occurs  through a crossover.

The ``instead-of-confinement'' regime heavily relies on  the presence of confined monopoles. One may wonder
how this phase can appear in the theory \eqref{fr4}. Classically the 't Hooft-Polyakov monopoles
are present in theories with the adjoint scalars which develop VEVs, as \ntwo SQCD. 
However, at the quantum level
the story becomes more subtle. It was shown that confined monopoles 
(seen as kinks in the CP$(N-1)$ model on the world sheet 
of the non-Abelian string) can survive the limit when the adjoint scalars are decoupled, see the  review paper
\cite{SYrev} and the recent publication \cite{IY}.

If we believe that SQCD gives us hints on the strong coupling dynamics in non-supersymmetric Yang-Mills theory,
we can arrive at a general conclusion that the phase transition ``perimeter law vs. area law"
is not the only characterization of confinement vs. non-confinement; other crossover transitions and phase transitions exist too.

\vspace{2mm}

% \begin{figure}[h]
%\centerline{\includegraphics[width=7cm]{levelcrosss}}
%\caption{\small A schematic picture of the crossover from large (on the left) to small 
%$v^2$ (on the right). Blue circles denote light quarks, while green circles denote
%light dual quarks. Monopole-antimonopole mesons are shown as in Fig.~\ref{figmeson}. The monopole-anti-monopole mesons on the left can be in the adjoint representation of the global group, in this case they decay in quarks and $W$ bosons. The  singlet component  with respect to the global group upon passing through the crossover and converts into dual quarks.}
%\label{figlevelcross}
%\end{figure}
 
%\begin{figure}
%\epsfxsize=6cm
%\centerline{levelcrosss}
%\caption{\small A schematic picture of the crossover from large (on the left) to small 
%$v^2$ (on the right). Blue circles denote light quarks, while green circles denote
%light dual quarks. Monopole-antimonopole mesons are shown as in Fig.~\ref{figmeson}. The monopole-anti-monopole mesons on the left can be in the adjoint representation of the global group, in this case they can decay in quarks. The  singlet component  with respect to the global group undergoes the crossover and converts into dual quarks.}
%\label{figlevelcross}
%\end{figure}
%

\section*{Acknowledgments}

This work  is supported in part by DOE grant DE-SC0011842. 
The work of A.Y. was  supported by William I. Fine Theoretical Physics Institute  of the  University 
of Minnesota, and by Russian State Grant for
Scientific Schools RSGSS-657512010.2. The work of A.Y. was supported by Russian Scientific Foundation 
under Grant No. 14-22-00281.


\newpage




\begin{thebibliography}{99}

{\small

  \bibitem{Frad} 
  E.~H.~Fradkin and S.~H.~Shenker,
{\em ``Phase Diagrams of Lattice Gauge Theories with Higgs Fields,}
  Phys.\ Rev.\ D {\bf 19}, 3682 (1979).
  %doi:10.1103/PhysRevD.19.3682
 % %%CITATION = doi:10.1103/PhysRevD.19.3682;%%
  %707 citations counted in INSPIRE as of 31 Dec 2016
  
\bibitem{SYrev}
M.~Shifman and A.~Yung,
Rev.\ Mod.\ Phys. {\bf 79}, 1139 (2007),
[arXiv:hep-th/0703267]; an expanded version in {\sl Supersymmetric Solitons,} 
(Cambridge University Press, 2009).
  %%CITATION = HEP-TH/0703267;%%
    
    \bibitem{16}
  M.~Shifman and A.~Yung,
{\em Lessons from supersymmetry: ``Instead-of-Confinement" Mechanism,}
  Int.\ J.\ Mod.\ Phys.\ A {\bf 29}, no. 27, 1430064 (2014),
 % doi:10.1142/S0217751X14300646
  [arXiv:1410.2900 [hep-th]].
  %%CITATION = doi:10.1142/S0217751X14300646;%%
  %4 citations counted in INSPIRE as of 14 Feb 2017
  
  \bibitem{SYdual}
  M.~Shifman and A.~Yung,
{\em Non-Abelian Duality and Confinement in ${\mathcal N}=2$  Supersymmetric QCD,}
  Phys.\ Rev.\  D {\bf 79}, 125012 (2009)
[arXiv:0904.1035 [hep-th]].
  %%CITATION = PHRVA,D79,125012;%% 
  
   \bibitem{15}
  M.~Shifman and A.~Yung,
{\em r Duality and `Instead-of-Confinement' Mechanism in ${\cal N}=1$ Supersymmetric QCD,}
  Phys.\ Rev.\ D {\bf 86}, 025001 (2012),
  %doi:10.1103/PhysRevD.86.025001
  [arXiv:1204.4165 [hep-th]].
  %%CITATION = doi:10.1103/PhysRevD.86.025001;%%
  %12 citations counted in INSPIRE as of 14 Feb 2017
  
    
   \bibitem{Bardakci} 
  K.~Bardakci and M.~B.~Halpern,
{\em Spontaneous breakdown and hadronic symmetries,}
  Phys.\ Rev.\ D {\bf 6}, 696 (1972).
 % doi:10.1103/PhysRevD.6.696
  %%CITATION = doi:10.1103/PhysRevD.6.696;%%
  %90 citations counted in INSPIRE as of 22 Nov 2016

  \bibitem{BPST} 
  A.~A.~Belavin, A.~M.~Polyakov, A.~S.~Schwartz and Y.~S.~Tyupkin,
{\em Pseudoparticle Solutions of the Yang-Mills Equations,}
  Phys.\ Lett.\  {\bf 59B}, 85 (1975).
%  doi:10.1016/0370-2693(75)90163-X
  %%CITATION = doi:10.1016/0370-2693(75)90163-X;%%
  %2601 citations counted in INSPIRE as of 31 Dec 2016
  
    \bibitem{hooft} 
  G.~'t Hooft,
 {\em Computation of the Quantum Effects Due to a Four-Dimensional Pseudoparticle,}
  Phys.\ Rev.\ D {\bf 14}, 3432 (1976)
  Erratum: [Phys.\ Rev.\ D {\bf 18}, 2199 (1978)].
  %doi:10.1103/PhysRevD.18.2199.3, 10.1103/PhysRevD.14.3432
  %%CITATION = doi:10.1103/PhysRevD.18.2199.3, 10.1103/PhysRevD.14.3432;%%
  %3714 citations counted in INSPIRE as of 31 Dec 2016
  
   
  \bibitem{renormalon}
G. 't Hooft,  {\em 	
Can We Make Sense Out of Quantum Chromodynamics?} in  {\sl The Whys Of Subnuclear Physics}, Erice
1977
ed. A. Zichichi (Plenum, New York, 1979), p. 943;
 B.~E.~Lautrup,
  %``On High Order Estimates in QED,''
  Phys.\ Lett.\ B {\bf 69}, 109 (1977);
  %%CITATION = PHLTA,B69,109;%%
  %192 citations counted in INSPIRE as of 03 Sep 2013;
G.~Parisi,
  %``Singularities of the Borel Transform in Renormalizable Theories,''
  Phys.\ Lett.\ B {\bf 76}, 65 (1978);
  %%CITATION = PHLTA,B76,65;%%
  %161 citations counted in INSPIRE as of 03 Sep 2013; 
%``On Infrared Divergences,''
  Nucl.\ Phys.\ B {\bf 150}, 163 (1979);
  %%CITATION = NUPHA,B150,163;%%
  %153 citations counted in INSPIRE as of 03 Sep 2013
 A.~H.~Mueller,
  %``On the Structure of Infrared Renormalons in Physical Processes at High-Energies,''
  Nucl.\ Phys.\ B {\bf 250}, 327 (1985).
  %%CITATION = NUPHA,B250,327;%%
  %278 citations counted in INSPIRE as of 03 Sep 2013
  
  \bibitem{ural}
  M.~Shifman,
{\em New and Old about Renormalons: in Memoriam Kolya Uraltsev,}
  Int.\ J.\ Mod.\ Phys.\ A {\bf 30}, no. 10, 1543001 (2015)
 % doi:10.1142/S0217751X15430010
  [arXiv:1310.1966 [hep-th]].
  %%CITATION = doi:10.1142/S0217751X15430010;%%
  %9 citations counted in INSPIRE as of 23 Jan 2017
  
    \bibitem{vene}
  G.~Veneziano,
{\em Some Aspects of a Unified Approach to Gauge, Dual and Gribov Theories,}
  Nucl.\ Phys.\ B {\bf 117}, 519 (1976).
  %doi:10.1016/0550-3213(76)90412-0
  %%CITATION = doi:10.1016/0550-3213(76)90412-0;%%
  %598 citations counted in INSPIRE as of 15 Feb 2017
  
      \bibitem{HSZ}
A. Hanany, M. Strassler, and A. Zaffaroni,
 {\em Confinement and strings in M{QCD},}
Nucl.\ Phys.\ B {\bf 513}, 87 (1998)
[hep-th/9707244].
%%CITATION = HEP-TH 9707244;%%

  \bibitem{VY}
A.~I.~Vainshtein and A.~Yung,
 {\em Type I superconductivity upon
 monopole condensation in Seiberg--Witten  theory,}
Nucl.\ Phys.\ B {\bf 614}, 3 (2001)
[hep-th/0012250].
%%CITATION = HEP-TH 0012250;%%

\bibitem{SYmon}
M.~Shifman and A.~Yung,
 {\em Non-Abelian string junctions as confined monopoles,}
Phys.\ Rev.\ D {\bf 70}, 045004 (2004)
[hep-th/0403149].
%%CITATION = HEP-TH 0403149;%%

\bibitem{HT1}
A.~Hanany and D.~Tong,
 {\em Vortices, instantons and branes,}
JHEP {\bf 0307}, 037 (2003)
[hep-th/0306150].
%%CITATION = HEP-TH 0306150;%%

\bibitem{ABEKY}
R.~Auzzi, S.~Bolognesi, J.~Evslin, K.~Konishi and A.~Yung,
 {\em Non-Abelian superconductors: Vortices and
  confinement in ${\mathcal N}=2$ 
  SQCD,}
Nucl.\ Phys.\ B {\bf 673}, 187 (2003)
[hep-th/0307287].
%%CITATION = HEP-TH 0307287;%%

\bibitem{HT2}
A. Hanany and D. Tong,
 {\em Vortex strings and four-dimensional gauge dynamics,}
JHEP {\bf 0404}, 066 (2004)
[hep-th/0403158].
%%CITATION = HEP-TH 0403158;%%

 \bibitem{SYtorkink}
M.~Shifman and A.~Yung,
{\em Non-Abelian Confinement in ${\mathcal N}=2$  Supersymmetric QCD: Duality and Kinks on
 Confining Strings,}
  Phys.\ Rev.\  D {\bf 81}, 085009 (2010)
  [arXiv:1002.0322 [hep-th]].
  %%CITATION = PHRVA,D81,085009;%%
  
  \bibitem{W79} 
E.~Witten,
{\em Instantons, the Quark Model, and the 1/N Expansion,}
  Nucl.\ Phys.\ B {\bf 149}, 285 (1979).
  %%CITATION = NUPHA,B149,285;%%
  %695 citations counted in INSPIRE as of 24 Feb 2015

 \bibitem{sph}
   R.~F.~Dashen, B.~Hasslacher and A.~Neveu,
{\em Nonperturbative Methods and Extended Hadron Models in Field Theory. 2. Two-Dimensional Models and Extended Hadrons,}
  Phys.\ Rev.\ D {\bf 10}, 4130 (1974);\\
%  doi:10.1103/PhysRevD.10.4130
  %%CITATION = doi:10.1103/PhysRevD.10.4130;%%
  %536 citations counted in INSPIRE as of 28 Jan 2017
  F.~R.~Klinkhamer and N.~S.~Manton,
{\em A Saddle Point Solution in the Weinberg-Salam Theory,}
  Phys.\ Rev.\ D {\bf 30}, 2212 (1984).
  %doi:10.1103/PhysRevD.30.2212
  %%CITATION = doi:10.1103/PhysRevD.30.2212;%%
  %1180 citations counted in INSPIRE as of 28 Jan 2017
  
  \bibitem{SW1}
N.~Seiberg and E.~Witten,
 {\em Electric - magnetic duality, monopole
 condensation, and confinement in ${\mathcal N}=2$ 
 supersymmetric Yang-Mills theory,}
Nucl. Phys. {\bf B426}, 19 (1994),
(E) {\bf B430},  485 (1994) [hep-th/9407087];
%%CITATION = NUPHA,B426,19;%%
%
% \bibitem{SW2}
%N.~Seiberg and E.~Witten,
 {\em Monopoles, duality and chiral symmetry breaking in ${\mathcal N}=2$ 
 supersymmetric QCD,}
Nucl. Phys. {\bf B431}, 484  (1994)
[hep-th/9408099].
%%CITATION = NUPHA,B431,484;%%

\bibitem{SW2}
N.~Seiberg and E.~Witten,
{\em Monopoles, duality and chiral symmetry breaking in ${\mathcal N}=2$ 
supersymmetric QCD,}
Nucl. Phys. {\bf B431}, 484  (1994)
[hep-th/9408099].
%%CITATION = NUPHA,B431,484;%%

  \bibitem{APS}
P.~Argyres, M.~Plesser and N.~Seiberg,
{\em The Moduli Space of ${\mathcal N}=2$  SUSY QCD and Duality in
${\mathcal N}=1$  SUSY QCD,}
Nucl. Phys. {\bf B471}, 159  (1996)
[hep-th/9603042].

\bibitem{ANO}
A.~Abrikosov, Sov.~Phys. JETP {\bf32}, 1442  (1957);
H.~Nielsen and P.~Olesen, Nucl.~Phys. {\bf B61}, 45 (1973).
[Reprinted in {\em Solitons and Particles}, Eds. C. Rebbi and G. Soliani
(World Scientific, Singapore, 1984), p. 365].

  \bibitem{BF}
A.~Bilal and F.~Ferrari,
 {\em The BPS spectra and superconformal points in massive ${\mathcal N}=2$  supersymmetric
   {QCD},}
  Nucl.\ Phys.\  B {\bf 516}, 175 (1998)
  [arXiv:hep-th/9706145].
  %%CITATION = NUPHA,B516,175;%%

  \bibitem{Cachazo2}
F.~Cachazo, N.~Seiberg and E.~Witten,
  {\em Chiral rings and phases of supersymmetric gauge theories,}
  JHEP {\bf 0304}, 018 (2003)
  [hep-th/0303207].
  %%CITATION = HEP-TH/0303207;%%
 
     
\bibitem{IY}
E.~Ievlev, A.~Yung,
{\em Non-Abelian strings in N=1 supersymmetric QCD}
arXiv:1704.03047.

  
   }
 



  





 
\end{thebibliography}
\end{document}